\documentclass[sigconf]{acmart}
\AtBeginDocument{%
  }

\copyrightyear{2025} 
\acmYear{2025} 
\setcopyright{acmlicensed}\acmConference[CCS '25]{Proceedings of the 2025 ACM SIGSAC Conference on Computer and Communications Security}{October 13--17, 2025}{Taipei, Taiwan}
\acmBooktitle{Proceedings of the 2025 ACM SIGSAC Conference on Computer and Communications Security (CCS '25), October 13--17, 2025, Taipei, Taiwan}
\acmDOI{10.1145/3719027.3765081}
\acmISBN{979-8-4007-1525-9/2025/10}

\acmSubmissionID{673}

\usepackage{tikz}
\usepackage{amsmath}
\usepackage{graphicx}
\usepackage{textcomp}
\usepackage{tabularx}
\newcolumntype{Y}{>{\centering\arraybackslash}X}

\newcolumntype{Z}{>{\hsize=1.3\hsize}X}
\newcolumntype{Q}{>{\hsize=.7\hsize}Y}
\newcolumntype{V}{>{\hsize=.15\hsize}X}
\usepackage{caption}
\usepackage{subcaption}
\usepackage{multirow}
\usepackage{multicol}
\usepackage{makecell}
\usepackage{xcolor}
\usepackage{xurl}
\usepackage{hyperref}
\usepackage{enumitem}
\usepackage{censor}
\usepackage{booktabs}
\usepackage[table]{xcolor} %
\usepackage{colortbl} %
\usepackage{geometry} %
\usepackage[most]{tcolorbox}
\usepackage{rotating}
\usepackage{wasysym}
\usepackage{balance}

\tcbset{
  takeawaystyle/.style={
    colback=white, %
    colframe=black, %
    fonttitle=\bfseries,%
    title={#1}, %
    attach boxed title to top center={yshift=-3mm}, %
    boxed title style={
      colback=black, %
      colframe=white, %
      boxrule=0mm, %
      boxsep=1mm, %
    },
    enhanced, %
    boxrule=0.5mm, %
    top=3mm, %
  },
}

\newcommand{\inquote}[1]{\textit{``#1''}}

\definecolor{yellow_a11y}{RGB}{242, 193, 68}
\definecolor{blue_a11y}{RGB}{27, 54, 93}

\begin{document}

\title{How Blind and Low-Vision Users Manage Their Passwords}

\author{Alexander Ponticello}
\orcid{0000-0001-6119-9701}
\affiliation{%
  \institution{CISPA Helmholtz Center for Information Security}
  \city{Saarbrücken}
  \country{Germany}
}
\email{alexander.ponticello@cispa.de}

\author{Filipo Sharevski}
\orcid{0000-0003-3058-7255}
\affiliation{%
  \institution{DePaul University}
  \city{Chicago}
  \state{IL}
  \country{USA}
}
\email{fsharevs@depaul.edu}

\author{Simon Anell}
\orcid{0009-0008-6358-832X}
\affiliation{%
  \institution{CISPA Helmholtz Center for Information Security}
  \city{Saarbrücken}
  \country{Germany}
}
\email{simon.anell@cispa.de}

\author{Katharina Krombholz}
\orcid{0000-0003-2425-3013}
\affiliation{%
  \institution{CISPA Helmholtz Center for Information Security}
  \city{Saarbrücken}
  \country{Germany}
}
\email{krombholz@cispa.de}

\renewcommand{\shortauthors}{Alexander Ponticello, Filipo Sharevski, Simon Anell, and Katharina Krombholz} 

\begin{abstract}
Managing passwords securely and conveniently is still an open problem for many users.
Existing research has examined users' password management strategies and identified pain points, such as security concerns, leading to insecure practices.
We investigate how Blind and Low-Vision (BLV) users tackle this problem and how password managers can assist them. 
This paper presents the results of a qualitative interview study with $N = 33$ BLV participants.
We found that all participants utilize password managers to some extent, which they perceive as fairly accessible.
However, the adoption is mainly driven by the convenience of storing and retrieving passwords. 
The security advantages -- generating strong, random passwords -- were avoided mainly due to the absence of \textit{practical} accessibility. 
Password managers do not adhere to BLV users' underlying needs for agency, which stem from experiences with inaccessible software and vendors who deprioritize accessibility issues. 
Underutilization of password managers leads BLV users to adopt insecure practices, such as reusing predictable passwords or resorting to ‘security through obscurity' by writing important credentials in braille.
We conclude our analysis by discussing the need to implement practical accessibility and usability improvements for password managers as a way of establishing trust and secure practices while maintaining BLV users' agency.

\end{abstract}

\begin{teaserfigure}
  \includegraphics[width=\textwidth]{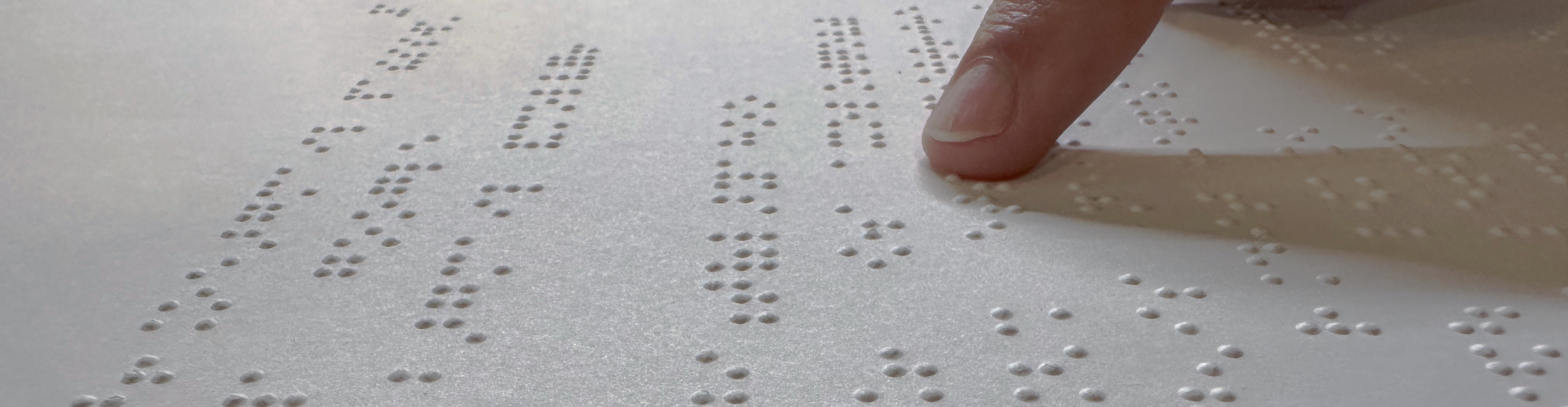}%
  \caption{A person is reading a password list written in braille.}%
  \Description{A white sheet of paper lies flat. It has several lines of braille characters printed on it. A person's finger is visible and placed on one of the lines to read.}%
  \label{fig:teaser}
\end{teaserfigure}



\maketitle

\section{Introduction} \label{sec:introduction}

Passwords are the most widespread security primitive, yet at the same time, they cause the most trouble. Insecure passwords usually contain predictable character sequences or combinations of characters, numbers, and symbols~\cite{nist-800-63b}. Therefore, people are asked to come up with long, complex, and close-to-random combinations. Since users often have hundreds of accounts~\cite{Viezelyte.2024}, each requiring a strong password, this results in an overwhelming number of complex combinations to generate, remember, and manage. 

This steep demand on people calls for automated password generation, management, and storage. Password Managers (PMs) are a common solution, as they can improve security and usability~\cite{nist-800-63b}. The improved security is achieved through the capability to generate unique, long, complex passwords. More usability is provided by relying on a single strong master password rather than memorizing different passwords for all accounts, and alleviating the need to type in passwords by automatically filling in credentials. For better convenience, people have the option to use standalone (e.g., LastPass, 1Password), browser-based (e.g., Chrome or Firefox), or Operating System (OS)-based PM applications (e.g., Apple Keychain). 

As PMs offer the opportunity to shed the persistent impression that people are the weakest link in the security chain, a line of research studied ``why people (don't) use password managers''~\cite{pearman-19-use-pw-manager, ray-21-pw-manager-old, mayer-22-pwm-edu}. In the initial inquiry, Pearman et al.~\cite{pearman-19-use-pw-manager} interviewed young adults to learn that people use OS-based PMs for convenience and standalone PMs for perceived better security. Those who did not use a PM cited a lack of awareness of the tool, a lack of motivation, and concerns about a single point of failure (vault breach or a master password compromise). The follow-up inquiry by Ray et al.~\cite{ray-21-pw-manager-old} focused explicitly on older adults, confirming the convenience and usability of not having to come up with, remember, and manage passwords, but also adding to the concerns about the cognitive decline associated with the aging process. Mayer et al.~\cite{mayer-22-pwm-edu} replicated the inquiry through a survey of a US university's faculty, staff, and students. While this follow-up study showed higher traction to the adoption of PMs, users often remained reluctant to use these tools due to a lack of trust and the burden of setup. 

In this paper, we expand this line of work by interviewing 33 people who have low vision or are blind\footnote{Legally blind with acuity of 20/200 or field-of-view of 20 degrees or less in the better eye with correction; low vision with acuity up to 20/70 and field-of-view larger than 20 degrees in the better eye with correction}.
Like sighted users, this population relies heavily on passwords. However, their use of PMs depends critically on the accessibility of these tools.
Existing evidence suggests that Blind and Low-Vision (BLV) users continue to face severe challenges due to insufficient accessibility support when entering passwords~\cite{Erinola2023-disabilityauth, alajarmeh2024password}. Hence, PMs present a natural opportunity to mitigate the disadvantages associated with insecure practices, without increasing users' exposure to security risks~\cite{napoli2021obstacles}. As this dimension of assisted usability seldom receives attention, we expanded the interview instrument from Pearman et al.~\cite{pearman-19-use-pw-manager} to include accessibility-specific questions. Our goal was to better understand how BLV individuals manage passwords and engage with PMs, our study is guided by the following research questions:

\smallskip

\begin{itemize}
    \item[\textbf{RQ1}] How do BLV users manage their passwords? \vspace{2pt}
    \item[\textbf{RQ2}] What role do PMs play? \vspace{2pt}
    \item[\textbf{RQ3}] How well do PMs align with BLV users' needs? 
\end{itemize}

\smallskip

We found that all participants employed PMs to some extend. Accessibility was an important factor in adopting such tools, and users navigated initial concerns through recommendations from BLV peers or an organic uptake. Perceived usefulness was the main motivation behind continuous usage, even though participants recognized security benefits. However, PM adoption was seldom holistic. We found that users often engaged in suboptimal practices (e.g., password reuse, predictable password generation). This behavior was influenced by several factors, most notably users’ perceived lack of autonomy and their aversion to relying on a system they experienced as unreliable.
Furthermore, our findings demonstrate that BLV users face similar issues when managing their passwords compared to sighted users, indicating that barriers exist beyond input mechanisms and across populations.
Additionally, we identified a number of challenges and mitigation strategies specific to BLV user, such as updates making PMs unusable by breaking assistive technology integration and people keeping physical password lists using braille, as depicted in Figure~\ref{fig:teaser}.

Based on our results, we derive implications for better accessible PMs. Most importantly, password generator should be adapted to BLV users need for autonomy (i.e., knowing what the password is), by producing passphrases that are easily legible by screen readers. Biometrics should be the primary form of authentication, as it works out-of-the-box for BLV users. PM vendors should also leverage testimonies from BLV users to help  alleviate accessibility concerns during the early stages of choosing a tool.

\section{Background and Related Work}\label{sec:background}

In this section, we present previous work on password management behavior. We emphasize the role of PMs, which often are the top recommendation to enhance users' security in the context of authentication by both experts~\cite{Gilbertson2025} and standardization bodies~\cite{nist-qb12}. Furthermore, we highlight advancements in providing accessible authentication to BLV users.

\subsection{Password Management Behavior}

Our work primarily extends a line of research initiated by Pearman et al.~\cite{pearman-19-use-pw-manager}. They studied users' password management behavior with a special focus on PMs. Their results show that users often engage in risky password behavior, such as password reuse. A lack of awareness about the technology and incomplete threat models were major factors for not using PMs. Among people who utilize PMs, driving factors for adoption diverged between convenience and security~\cite{pearman-19-use-pw-manager, alkaldi-16-adoption-phone-managers, Munyendo2023switchingpassmngrs}. These factors also primarily influenced the choice of PM. Usability was crucial to reduce frustration and expand adoption. Ray et al.~\cite{ray-21-pw-manager-old} replicated this study with senior citizens. This population had overlapping issues with younger adults, but some were distinct, e.g., higher distrust in cloud storage. Likewise, older adults placed greater emphasis on recommendations from friends and family. Mayer et al.~\cite{mayer-22-pwm-edu} further replicated these studies in a large US university context. Focusing on awareness, password behavior, and PM adoption strategies, they showed that adoption had increased compared to previously reported figures~\cite{couillard2015password}. They highlight how institutions can foster more secure password behavior by facilitating PM access. 
Research of PM usage in the wild~\cite{oesch-22-started-pw-manager} 
confirmed these findings while showcasing operational issues of current PMs, such as entering randomly generated passwords into new systems, which negatively impacted profound adoption.

Research looked into barriers presenting themselves at the initial stages of adopting a PM. Lack of awareness and misunderstood security benefits, attributed to poor advertisement, negatively impacted users' motivation~\cite{alkaldi-16-adoption-phone-managers}. Usability issues and distrust in the technology raised the entry barrier too far for large groups of users~\cite{amft-23-cred-management}.
Missing guidance also fostered insecure behavior in the process of switching PMs~\cite{Munyendo2023switchingpassmngrs}.
Meanwhile, Alkaldi and Renaud~\cite{alkaldi_encouraging_2019} showed the beneficial impact of catering to the users' need for self-determination~\cite{ryan_self-determination_2000} on PM adoption.

Randomly generated passwords are crucial in defending against guessing attacks~\cite{wang-16-online-guessing} while preventing password reuse~\cite{stobert-14-pw-life-cycle}. Therefore, it is vital to steer users into utilizing this functionality to maximize security gains~\cite{zibaei-22-random}. Adopting this feature was sparse and mainly limited to low-value accounts where convenience out-valued security~\cite{Zibaei2023simple-nudgepassmgnrs, amft-23-cred-management}. Important accounts were often left out due to distrust in PM vendors~\cite{amft-23-cred-management}.
Further contributing to PMs' security are credential auditing tools, which inform users about weak or breached passwords~\cite{Hutchinson2024-password-audit}. While the benefits are clear, there is little consistency in the implementation across systems, which confuses users~\cite{Hutchinson2024-password-audit}. Furthermore, some systems lacked compatibility with PMs, e.g., by not accepting randomly generated passwords~\cite{hutchinson2024measuring}. The situation was especially dire on mobile phones, where PMs suffered extensive usability issues~\cite{Seiler-Hwang2019-dontusepassmngrs}.

Our research expands this line of research by including the experiences of BLV users. Their unique way of interacting with systems via assistive technologies creates previously unstudied circumstances for authentication and PMs, e.g., entering a password using a screen reader in the presence of bystanders. At the same time, BLV users can have heightened incentives for adopting a PM, such as having access to the much preferred biometric authentication~\cite{jain_smartphone_2021,Faustino2027understanding} on more platforms.

\subsection{Accessibility of Password Management}

Previous work also investigated BLV users' experiences around authentication. Faustino et al.~\cite{Faustino2027understanding} studied how BLV users unlock their smartphone. Erinola et al.~\cite{Erinola2023-disabilityauth} explored challenges faced by people with disabilities around various authentication methods. Both studies highlight that no method is fully accessible, and users are forced to adopt workarounds that expose them to additional threats. The studies emphasize the ample potential of PMs to enhance BLV users' security.
However, BLV users held concerns regarding these tools' accessibility~\cite{napoli2021obstacles}.

Previous work explored better-suited authentication methods for BLV users, as existing systems often showed fundamental accessibility issues, such as items not reachable by screen readers or too short time limits for confirming actions~\cite{Barbosa2016UniPass}.
Briotto Faustino et al.~\cite{faustino2020bend} proposed a bendable authentication method to replace PINs on mobile phones. 
Wolf et al.~\cite{wolf_perceptions_2017} evaluated a tactile aid for mobile authentication, which could assist in scenarios where bystanders were present and conventional assistive technology could undermine security.
Finally, Barbosa et al.~\cite{Barbosa2016UniPass} developed a desktop-based PM specifically for BLV users. It utilizes smart devices to provide users with biometric authentication instead of a master password. Comparing the prototype to existing solutions suggested that biometric authentication enhances users' perceived security. Alajarmeh et al.~\cite{alajarmeh2024password} investigated the awareness and usage of PMs among BLV users. The quantitative results of their survey show that BLV users have a good awareness of PMs, with 62\% reporting practical experiences. Active usage, however, was still low at 37\%.

Our work expands existing literature by in-depth investigating BLV users' password behavior and perceptions of PMs using a qualitative approach. Our findings are vital to identifying underlying barriers to adoption and perceptions leading to suboptimal password behavior.

\section{Study Methodology} \label{sec:study}
Previous work established the importance of investigating password management behavior to better understand how PMs can assist users more effectively~\cite{pearman-19-use-pw-manager,ray-21-pw-manager-old,mayer-22-pwm-edu}. We extend the literature by exploring how BLV users maintain their credentials. With the ultimate goal of removing barriers to accessible, secure, and usable authentication for everyone.

\subsection{Recruitment}

\paragraph{Protocol.} We received IRB approval to recruit participants who are blind or have low vision, are from the US, are 18 years or older, can understand and converse in English, and use passwords for online authentication. As one of the researchers had access to a diverse pool of BLV individuals, we recruited potential participants through these contacts. We used a formal email approved by our IRB to approach each potential participant containing a copy of the consent form for collecting and processing their data. Participants could then sign up for predefined time slots and receive a Zoom link for the study interview. We also employed snowball sampling, where we asked participants about any BLV individuals they knew who could be interested in participating in this study. This became especially relevant after we learned that several participants started using PMs on the recommendation of fellow BLV peers. In those cases, we asked them to refer our study to them. Participants also took it upon themselves to advertise our study in their local BLV community.

Based on previous work, which involved 30~\cite{pearman-19-use-pw-manager} and 26~\cite{ray-21-pw-manager-old} participants, respectively, we expected to recruit around 30 individuals for our study. Since our second stage recruitment efforts (i.e., snowball sampling, advertising study in online communities) resulted in some latency before participants signed up for a study slot, we overshot our target of 30 participants. However, we decided not to turn away participants, considering how that might make them more reluctant to participate in future studies. The study took around 60 minutes, and we compensated each participant with a \$60 Amazon eGift card. Table~\ref{tab:demographics} shows the demographic data.

\begin{table}[t!]
    \newlength\atroxfactor %
    \setlength\atroxfactor{0.1mm} %
    \newcommand{\atroxbar}[2]{\raisebox{0.2mm}{\fboxsep0pt \fboxrule0pt\fcolorbox{#1}{#1}{\vphantom{x}\hspace{#2\atroxfactor}\null}}}
    \colorlet{b5}{blue!60!white} %
    \colorlet{b1}{red!45!green}
    \colorlet{b2}{blue!40!green}
    \colorlet{b3}{red!40!white}
    \colorlet{b4}{red!40!yellow}
    \centering
    \caption{Participant demographics. Total $N=33$. Participants often used multiple assistive technologies in parallel.}
    \begin{tabular}{m{0.26\textwidth} | r  l}
        \hline
         \textbf{Demographics} & \multicolumn{2}{l}{\textbf{Participants}}  \\
         \hline
         \textbf{Gender} \\
         Female & 17  & \atroxbar{b5}{52}\\
         Male & 15  & \atroxbar{b5}{46}\\
         Non-Binary & 1  & \atroxbar{b5}{3}\\
         \hline
         \textbf{Race/Ethnicity} \\
         White & 21  & \atroxbar{b1}{63.5}\\
         Hispanic & 3  & \atroxbar{b1}{9}\\
         Asian & 3 &  \atroxbar{b1}{9}\\ %
         Middle Eastern & 2  & \atroxbar{b1}{6}\\
         Black/African American & 2 & \atroxbar{b1}{6}\\
         Multiracial & 2 & \atroxbar{b1}{6}\\
         \hline
         \textbf{Age} \\
    	 18 - 28 & 6  & \atroxbar{b2}{18}\\
	     29 - 39 & 13 & \atroxbar{b2}{39}\\
	     40 - 50 & 11& \atroxbar{b2}{33}\\
	     51+ & 3 &\atroxbar{b2}{9}\\
         \hline
         \textbf{Highest Education} \\
	     High School & 2  & \atroxbar{b3}{6}\\
          Some College & 10 & \atroxbar{b3}{30.30}\\ %
          Associate’s Degree & 2  & \atroxbar{b3}{6}\\
          Bachelor’s & 7 &  \atroxbar{b3}{21.21}\\ 
          Master’s & 8 & \atroxbar{b3}{24.24}\\ %
	     Gradschool & 2 &  \atroxbar{b3}{6.1}\\
	     PhD & 2 & \atroxbar{b3}{6.1}\\
	    \hline
         \textbf{Assistive Technology} \\
         VoiceOver screen reader & 31 & \atroxbar{b4}{93.93}\\
         JAWS screen reader & 27 &  \atroxbar{b4}{81.82}\\
         NVDA screen reader & 11 &  \atroxbar{b4}{33.3}\\
         Braille display & 7 & \atroxbar{b4}{21.21}\\
         Screen magnifier & 5  & \atroxbar{b4}{15.15}\\ %
         TalkBack screen reader & 3  & \atroxbar{b4}{9.09}\\
         Large text & 3  & \atroxbar{b4}{9.09}\\
         Smart glasses & 2 & \atroxbar{b4}{6.06}\\
         Narrator screen reader & 1 &  \atroxbar{b4}{3.03}\\
         Seeing AI image-to-speech system & 1 &  \atroxbar{b4}{3.03}\\
         Telemicroscopes& 1 &  \atroxbar{b4}{3.03}\\
         High contrast& 1  & \atroxbar{b4}{3.03}\\
         \hline
    \end{tabular}
    \label{tab:demographics}
\end{table}

\paragraph{Sample size.} We did not use the concept of \textit{saturation}~\cite{glaser_discovery_2017} following criticism of applying this concept to non-grounded theory qualitative research~\cite{oreilly_unsatisfactory_2013, braun2021thematic}. Instead, we turn to the concept of \textit{information power} as proposed by Malterud et al.~\cite{malterud_sample_2016}, which allows us to evaluate our sample size along a five-axis model: aim, specificity, use of theory, dialog, analysis. The \textit{aim} of our study was at the more narrow side of the spectrum, given that prior research has explored password management of a general populations~\cite{pearman-19-use-pw-manager}, older adults~\cite{ray-21-pw-manager-old}, or professional users~\cite{mayer-22-pwm-edu}. Our research questions are bound by the BLV population that depends on assistive technology for any password-related task. The \textit{specificity} of our study is dense, as all participants were technologically literate, ideally positioned to share lived experiences with passwords and PMs they have used, are currently using, or might use in the future. 

Our final analysis relies on very little pre-existing \textit{theory}, i.e., it is exploratory in nature. The quality of the \textit{dialog} of our interviews is medium to strong. The involved researchers had extensive experience conducting interviews. However, participants might have been inclined to adapt to the fact that the interviewers were sighted. Hence, they spent more time explaining their lived experiences to outsiders than going into depth. Finally, our \textit{analysis} was cross-cases, as our goal was to capture diverse password management and password manager usage experiences. Given these factors, we deem our study's \textit{information power} sufficiently strong. Reflecting on the thematic analysis of our data, a slightly smaller sample (20-25 participants) might have provided a similarly sufficient information power level, primarily if the research questions explicitly focused on screen-reader users, which was the predominant assistive technology in our sample (a limitation we acknowledge). 

\subsection{Interviews}
\paragraph{Instrument.} We based our semi-structured interview script on the one established by Pearman et al., which investigated \textit{``adoption and effective use of password managers and password-generation features''}~\cite{pearman-19-use-pw-manager}. The same script was successfully reused with a similar purpose and involving an at-risk user population (older adults)~\cite{ray-21-pw-manager-old}. It includes questions about general password usage, potential experiences with PMs, and reasons for (non-)adoption. We expanded the existing interview script to include accessibility topics. The full interview script is available on request.

We scheduled each interview for 60 minutes. Two researchers were present during every interview slot. One primary researcher conducted the interviews, and one senior researcher observed and occasionally asked follow-up questions. We clarified the purpose of the study as early as recruiting potential participants in the initial recruitment email. We invited potential participants to an audio-only Zoom interview, which allowed them to answer our questions from the safety and intimacy of their own homes, which we regarded especially important for BLV individuals. We informed the participants that they did not need to turn on their webcam while the primary interviewer transmitted their video. At the start of each session, we obtained consent to record the audio, which subsequently was automatically transcribed by Zoom. Immediately after the interview, we downloaded the Zoom transcripts (no later than 30 minutes after the conclusion and closing of the session), transferred them into text documents, and anonymized the resulting transcripts by removing every mention of the participant's name or the name of any acquaintances. After this, we deleted the audio records.

At the end of each session, we allowed the participants to ask questions about our research. We also clarified any misconceptions they might have had about passwords and PMs. We briefed the participants on certain features of their PM, which we thought could be helpful to them based on their descriptions. For instance, we explained to a participant, who expressed frustration about having to type in random character sequences, how to generate intelligible passphrases instead. The two researchers involved in the interviewing took notes throughout the process, in addition to the automatic Zoom transcript, and routinely reflected on the procedure and the insights they gained.
In some cases, we identified interesting topics that participants brought up and decided to adjust the follow-up questions to probe deeper into these issues in subsequent interviews. The most prominent example of this was noting down passwords in braille. This allowed us to explore directions not yet covered by our script.

\paragraph{Pilot testing.} After reviewing our interview setup with all team members, we decided to conduct pilot interviews with actual study participants. We wanted to make sure that the adaptations of the interview script worked, too, given that the main part was already tried and tested in previous works with other populations. After running the interviews with two participants, the two researchers involved determined no need for adjustments; therefore, we included the pilot data in our main analysis.

\subsection{Analysis}
The basis of our analysis was the transcripts of our interviews and the notes we took during the data collection. Our initial intention was to analyze the data analogously to previous work on PM usage~\cite{pearman-19-use-pw-manager,ray-21-pw-manager-old}, which also served as the foundation for our interview script. We planned on reusing their codebook while adding new codes for topics related to accessibility and BLV users' experiences. Doing so would allow us to compare our results to previous studies with other populations (i.e., sighted and older adults). During the initial round of deductive coding, it became apparent to us that such a comparison to the aforementioned works and the population they studied would be inappropriate due to the inherent heterogeneity, not only in the users' interaction mechanisms but also in their everyday reality --besides, none of the previous populations depends on an intermediary, such as a screen reader or magnifier, to access passwords and PMs. 

Examples include a heightened vulnerability (e.g., when authenticating while in public) and a different notion of trust (e.g., relying on sighted allies such as family members for specific password tasks). Therefore, the entire research team agreed to abandon this initial approach and shift to an inductive, exploratory approach without attempting to fit our data into any existing theory. Recognizing that all research team members are sighted, we acknowledged our inability to personally experience our participants' reality as \textit{insiders}. Instead, our understanding is informed by second-hand experience through conversations with BLV individuals as \textit{outsiders}. This positioning resonates with the foundational concepts of the reflexive thematic analysis, as described by Braun and Clarke~\cite{braun2021thematic}, which we chose as our analytical approach. It allowed us to explore our data freely, but the reflexive nature of this approach was crucial, as it enabled us to scrutinize any preconceptions we held.

The primary researcher and a second junior researcher conducted the central part of the reflexive thematic analysis.
While analyzing the data, both researchers kept a reflexive diary to critically examine their thoughts. Following the phases of thematic analysis~\cite{braun-06-thematic-analysis}, we began by familiarizing ourselves with the data by reading through transcripts on a per-interview and per-question basis. Afterward, the two researchers divided the interviews evenly (odd and even numbered transcripts), taking into account that the focus of the interviews had shifted slightly throughout the study. Each researcher inductively coded their data set. They met after every four to six interviews to discuss their coding. At this stage, they did not exchange or merge codebooks but instead pitched ideas to each other. After this initial round, each interview had been coded by one researcher. Then, the two researchers exchanged their coded data sets, examining the newly received transcripts. In this step, they sparsely assigned missing codes and focused on taking notes towards generating initial themes. 

Finally, both researchers met for consecutive days to refine and finalize the themes. During these meetings, they first immersed ourselves in the data, using over 300 sticky notes with codes and annotations. Then, they continuously checked whether the themes aligned with what they heard during the interviews. Eventually, they discussed the final themes with the senior researcher present during the interviews and helped review the thematic analysis and come up with names for each of them. They wrote the results around the themes, selecting quotations to evidence the claims and enable others to independently assess the alignment between the data and their understanding and interpretation of them~\cite{braun2021thematic}.

\subsection{Positionality}

All authors are from a Western, educated, industrialized, rich, and democratic (WEIRD) background. Three reside in Germany and one in the US. Three identify as male and one as female. None of the authors is legally blind or has low vision. The primary author is a usable security researcher and has previous experience conducting qualitative studies and working with BLV users. He became interested in the unique security and privacy challenges this population faces after talking to acquainted blind individuals. He familiarized himself with assistive technologies for BLV users prior to the start of the project. One senior researcher has similar extensive experiences from previous studies. His allyship and advocacy for removing security and privacy barriers for BLV individuals enabled the research team to access a diverse pool of potential participants. 

This relationship helped establish trust between the sighted researchers and the BLV participants, emphasizing the participatory principle of ``\textit{nothing about us without us}'' for their involvement~\cite{jurgens2005nothing}. The primary and the senior author jointly conducted all interviews. All members of the research team share a desire to make security technology more accessible and include at-risk populations in the scientific discourse. Initially, the primary researcher was concerned about conducting research with a population they were not part of. Our participants helped dissipate these concerns by expressing their pleasure in contributing and describing the study as important for improving accessible authentication. We further addressed these concerns by having an external researcher who is blind review our study protocol and approach beforehand, as well as our analysis and interpretations afterward.

\section{Findings} \label{sec:results}

This section presents the final themes we agreed upon after engaging in a reflexive thematic analysis process. Our findings comprise two main themes:
\textbf{(I) The presence of \textit{technical} accessibility enables adoption of PMs}.
All participants use PMs to some extend. They overcame initial challenges with accessibility either through recommendations or by gradually finding solutions on their own. Perceived accessibility goes hand in hand with improved usefulness. Therefore, users value PMs primarily for their convenience rather than their security benefits.
\textbf{(II) The absence of \textit{practical} accessibility restricts secure password behavior}.
In contrast, we found that, although PMs are widely used, barriers prevent holistic adoption. These barriers drive users towards suboptimal password behavior. The absence of \textit{practical} accessibility in this context is rooted in a lack of autonomy and a dependence on fragile systems, which current PMs impose on BLV users.

\subsection{Theme I: The Presence of \textit{Technical} Accessibility Enables Adoption of PMs}

After analyzing our data, we determined that PMs were central to the participants' password behavior. All had used a PM at some point, at least to some extent. The users' experiences differ in how they initially started using a PM, which specific software they used, and which features they employed. Table~\ref{tab:password_managers} details the various applications participants utilized and the corresponding uptake strategy. Before digging deeper into how PMs integrate into BLV users' everyday digital lives, we describe both approaches.

\begin{table*}[t]
\caption{Overview of PMs used by participants, grouped by type: OS-integrated, browser-integrated, and third-party standalone tools. The shading of each cell indicates whether the user’s uptake of that PM was deliberate or impromptu (see Section~\ref{sec:uptake}).}
\label{tab:password_managers}
\begin{tabular}{|p{1.5cm}|*{33}{>{\centering\arraybackslash}p{0.115cm}|}}
\hline
\textit{Participant}
& \makebox[0.12cm][c]{1}  & \makebox[0.12cm][c]{2}  & \makebox[0.12cm][c]{3}  & \makebox[0.12cm][c]{4}  & \makebox[0.12cm][c]{5}
& \makebox[0.12cm][c]{6}  & \makebox[0.12cm][c]{7}  & \makebox[0.12cm][c]{8}  & \makebox[0.12cm][c]{9}  & \makebox[0.12cm][c]{10}
& \makebox[0.12cm][c]{11} & \makebox[0.12cm][c]{12} & \makebox[0.12cm][c]{13} & \makebox[0.12cm][c]{14} & \makebox[0.12cm][c]{15}
& \makebox[0.12cm][c]{16} & \makebox[0.12cm][c]{17} & \makebox[0.12cm][c]{18} & \makebox[0.12cm][c]{19} & \makebox[0.12cm][c]{20}
& \makebox[0.12cm][c]{21} & \makebox[0.12cm][c]{22} & \makebox[0.12cm][c]{23} & \makebox[0.12cm][c]{24} & \makebox[0.12cm][c]{25}
& \makebox[0.12cm][c]{26} & \makebox[0.12cm][c]{27} & \makebox[0.12cm][c]{28} & \makebox[0.12cm][c]{29} & \makebox[0.12cm][c]{30}
& \makebox[0.12cm][c]{31} & \makebox[0.12cm][c]{32} & \makebox[0.12cm][c]{33} \\
\hline 
Keychain & \cellcolor{blue_a11y}{\tiny \textcolor{blue_a11y}{d}} & \cellcolor{yellow_a11y}{\tiny \textcolor{yellow_a11y}{i}} & \cellcolor{yellow_a11y}{\tiny \textcolor{yellow_a11y}{i}} & \cellcolor{yellow_a11y}{\tiny \textcolor{yellow_a11y}{i}} & \cellcolor{yellow_a11y}{\tiny \textcolor{yellow_a11y}{i}} & \cellcolor{yellow_a11y}{\tiny \textcolor{yellow_a11y}{i}} &  {\tiny \textcolor{white}{no}}   & \cellcolor{yellow_a11y}{\tiny \textcolor{yellow_a11y}{i}} & \cellcolor{blue_a11y}{\tiny \textcolor{blue_a11y}{d}} & \cellcolor{yellow_a11y}{\tiny \textcolor{yellow_a11y}{i}} & \cellcolor{yellow_a11y}{\tiny \textcolor{yellow_a11y}{i}} & \cellcolor{blue_a11y}{\tiny \textcolor{blue_a11y}{d}} &   {\tiny \textcolor{white}{no}}   & \cellcolor{yellow_a11y}{\tiny \textcolor{yellow_a11y}{i}} &  {\tiny \textcolor{white}{no}}    & \cellcolor{yellow_a11y}{\tiny \textcolor{yellow_a11y}{i}} & \cellcolor{blue_a11y}{\tiny \textcolor{blue_a11y}{d}} & \cellcolor{yellow_a11y}{\tiny \textcolor{yellow_a11y}{i}} & \cellcolor{yellow_a11y}{\tiny \textcolor{yellow_a11y}{i}} & \cellcolor{yellow_a11y}{\tiny \textcolor{yellow_a11y}{i}} & \cellcolor{yellow_a11y}{\tiny \textcolor{yellow_a11y}{i}} &  {\tiny \textcolor{white}{no}}    &  {\tiny \textcolor{white}{no}}    & \cellcolor{blue_a11y}{\tiny \textcolor{blue_a11y}{d}} & \cellcolor{yellow_a11y}{\tiny \textcolor{yellow_a11y}{i}} &  {\tiny \textcolor{white}{no}}    & \cellcolor{blue_a11y}{\tiny \textcolor{blue_a11y}{d}} & \cellcolor{yellow_a11y}{\tiny \textcolor{yellow_a11y}{i}} & \cellcolor{yellow_a11y}{\tiny \textcolor{yellow_a11y}{i}} & {\tiny \textcolor{white}{no}}  & \cellcolor{yellow_a11y}{\tiny \textcolor{yellow_a11y}{i}}    &  {\tiny \textcolor{white}{no}} &\cellcolor{yellow_a11y}{\tiny \textcolor{yellow_a11y}{i}}\\\hline \hline
Chrome   &  {\tiny \textcolor{white}{no}}    &  {\tiny \textcolor{white}{no}}    & {\tiny \textcolor{white}{no}}  &   {\tiny \textcolor{white}{no}}   &   {\tiny \textcolor{white}{no}}   & \cellcolor{blue_a11y}{\tiny \textcolor{blue_a11y}{d}} & \cellcolor{yellow_a11y}{\tiny \textcolor{yellow_a11y}{i}} &  {\tiny \textcolor{white}{no}}    & \cellcolor{blue_a11y}{\tiny \textcolor{blue_a11y}{d}} &  {\tiny \textcolor{white}{no}}    & \cellcolor{yellow_a11y}{\tiny \textcolor{yellow_a11y}{i}} & \cellcolor{yellow_a11y}{\tiny \textcolor{yellow_a11y}{i}} & \cellcolor{blue_a11y}{\tiny \textcolor{blue_a11y}{d}} &  {\tiny \textcolor{white}{no}}    &   {\tiny \textcolor{white}{no}}   & \cellcolor{yellow_a11y}{\tiny \textcolor{yellow_a11y}{i}} &  {\tiny \textcolor{white}{no}}    & \cellcolor{yellow_a11y}{\tiny \textcolor{yellow_a11y}{i}} & \cellcolor{yellow_a11y}{\tiny \textcolor{yellow_a11y}{i}} &     & \cellcolor{yellow_a11y}{\tiny \textcolor{yellow_a11y}{i}} & & \cellcolor{yellow_a11y}{\tiny \textcolor{yellow_a11y}{i}} & {\tiny \textcolor{white}{no}}  &  {\tiny \textcolor{white}{no}}    & \cellcolor{yellow_a11y}{\tiny \textcolor{yellow_a11y}{i}} &   {\tiny \textcolor{white}{no}}   &  {\tiny \textcolor{white}{no}}    & \cellcolor{yellow_a11y}{\tiny \textcolor{yellow_a11y}{i}} &  {\tiny \textcolor{white}{no}}    & \cellcolor{yellow_a11y}{\tiny \textcolor{yellow_a11y}{i}} &   {\tiny \textcolor{white}{no}}   &  {\tiny \textcolor{white}{no}}    \\\hline
Firefox  &  {\tiny \textcolor{white}{no}}    &   {\tiny \textcolor{white}{no}}   &   {\tiny \textcolor{white}{no}}   &  {\tiny \textcolor{white}{no}}    &  {\tiny \textcolor{white}{no}}    &   {\tiny \textcolor{white}{no}}   &  {\tiny \textcolor{white}{no}}    &  {\tiny \textcolor{white}{no}}    &  {\tiny \textcolor{white}{no}}    &  {\tiny \textcolor{white}{no}}    &  {\tiny \textcolor{white}{no}}    &   {\tiny \textcolor{white}{no}}   &  {\tiny \textcolor{white}{no}}    &  {\tiny \textcolor{white}{no}}    &  {\tiny \textcolor{white}{no}}    & \cellcolor{yellow_a11y}{\tiny \textcolor{yellow_a11y}{i}} & {\tiny \textcolor{white}{no}}     &  {\tiny \textcolor{white}{no}}    &   {\tiny \textcolor{white}{no}}   &  {\tiny \textcolor{white}{no}}    & \cellcolor{yellow_a11y}{\tiny \textcolor{yellow_a11y}{i}} &    {\tiny \textcolor{white}{no}}  &  {\tiny \textcolor{white}{no}}    &  {\tiny \textcolor{white}{no}}    & {\tiny \textcolor{white}{no}}    &  {\tiny \textcolor{white}{no}}    &  {\tiny \textcolor{white}{no}}    &  {\tiny \textcolor{white}{no}}    &   {\tiny \textcolor{white}{no}}   &  {\tiny \textcolor{white}{no}}    &  {\tiny \textcolor{white}{no}}    &  {\tiny \textcolor{white}{no}}    &   {\tiny \textcolor{white}{no}}   \\\hline
Edge  &  {\tiny \textcolor{white}{no}}    &   {\tiny \textcolor{white}{no}}   &  {\tiny \textcolor{white}{no}}    &   {\tiny \textcolor{white}{no}}   & {\tiny \textcolor{white}{no}}  &   {\tiny \textcolor{white}{no}}   &   {\tiny \textcolor{white}{no}}   &   {\tiny \textcolor{white}{no}}   &   {\tiny \textcolor{white}{no}}   &  {\tiny \textcolor{white}{no}}    &   {\tiny \textcolor{white}{no}}   &  {\tiny \textcolor{white}{no}}    &  {\tiny \textcolor{white}{no}}    &  {\tiny \textcolor{white}{no}}    &  {\tiny \textcolor{white}{no}}    &  {\tiny \textcolor{white}{no}}    &  {\tiny \textcolor{white}{no}}    &  {\tiny \textcolor{white}{no}}    &  {\tiny \textcolor{white}{no}}    &  {\tiny \textcolor{white}{no}}    &  {\tiny \textcolor{white}{no}}    &  {\tiny \textcolor{white}{no}}    &  {\tiny \textcolor{white}{no}}    &   {\tiny \textcolor{white}{no}}   &  \cellcolor{yellow_a11y}{\tiny \textcolor{yellow_a11y}{i}}    &  {\tiny \textcolor{white}{no}}    &  {\tiny \textcolor{white}{no}}    &  {\tiny \textcolor{white}{no}}    &  {\tiny \textcolor{white}{no}}    &  {\tiny \textcolor{white}{no}}    &  {\tiny \textcolor{white}{no}}    &  {\tiny \textcolor{white}{no}}    &  \cellcolor{yellow_a11y}{\tiny \textcolor{yellow_a11y}{i}}   \\\hline\hline
1Password  & \cellcolor{blue_a11y}{\tiny \textcolor{blue_a11y}{d}} & \cellcolor{blue_a11y}{\tiny \textcolor{blue_a11y}{d}} &  {\tiny \textcolor{white}{no}}    & \cellcolor{blue_a11y}{\tiny \textcolor{blue_a11y}{d}} &  {\tiny \textcolor{white}{no}}    & \cellcolor{blue_a11y}{\tiny \textcolor{blue_a11y}{d}} &  {\tiny \textcolor{white}{no}}    &  {\tiny \textcolor{white}{no}}    &  {\tiny \textcolor{white}{no}}    &  {\tiny \textcolor{white}{no}}    &  {\tiny \textcolor{white}{no}}    &  {\tiny \textcolor{white}{no}}    & \cellcolor{blue_a11y}{\tiny \textcolor{blue_a11y}{d}} & \cellcolor{blue_a11y}{\tiny \textcolor{blue_a11y}{d}} & \cellcolor{blue_a11y}{\tiny \textcolor{blue_a11y}{d}} &  {\tiny \textcolor{white}{no}}    &  {\tiny \textcolor{white}{no}}    &   {\tiny \textcolor{white}{no}}   &  {\tiny \textcolor{white}{no}}    &  {\tiny \textcolor{white}{no}}    &  {\tiny \textcolor{white}{no}}    & \cellcolor{blue_a11y}{\tiny \textcolor{blue_a11y}{d}} & \cellcolor{blue_a11y}{\tiny \textcolor{blue_a11y}{d}} & \cellcolor{blue_a11y}{\tiny \textcolor{blue_a11y}{d}} &  {\tiny \textcolor{white}{no}}    &  {\tiny \textcolor{white}{no}}    & \cellcolor{blue_a11y}{\tiny \textcolor{blue_a11y}{d}} &  {\tiny \textcolor{white}{no}}    &   {\tiny \textcolor{white}{no}}   & {\tiny \textcolor{white}{no}}  &   {\tiny \textcolor{white}{no}}   & {\tiny \textcolor{white}{no}}  & {\tiny \textcolor{white}{no}}  \\\hline
KeePass  &  \cellcolor{blue_a11y}{\tiny \textcolor{blue_a11y}{d}}   &   {\tiny \textcolor{white}{no}}   &  {\tiny \textcolor{white}{no}}    &   {\tiny \textcolor{white}{no}}   & {\tiny \textcolor{white}{no}}  &  {\tiny \textcolor{white}{no}}    &  {\tiny \textcolor{white}{no}}    &  {\tiny \textcolor{white}{no}}    &  {\tiny \textcolor{white}{no}}    &  {\tiny \textcolor{white}{no}}    &   {\tiny \textcolor{white}{no}}   &  {\tiny \textcolor{white}{no}}    &  {\tiny \textcolor{white}{no}}    &   {\tiny \textcolor{white}{no}}   &  {\tiny \textcolor{white}{no}}    &  {\tiny \textcolor{white}{no}}    &  {\tiny \textcolor{white}{no}}    &   {\tiny \textcolor{white}{no}}   &  {\tiny \textcolor{white}{no}}    &  {\tiny \textcolor{white}{no}}    &  {\tiny \textcolor{white}{no}}    &  {\tiny \textcolor{white}{no}}    &  {\tiny \textcolor{white}{no}}    &  {\tiny \textcolor{white}{no}}    &  {\tiny \textcolor{white}{no}}    &  {\tiny \textcolor{white}{no}}    &   {\tiny \textcolor{white}{no}}   &   {\tiny \textcolor{white}{no}}   &  {\tiny \textcolor{white}{no}}    &  {\tiny \textcolor{white}{no}}    &   {\tiny \textcolor{white}{no}}   &  \cellcolor{blue_a11y}{\tiny \textcolor{blue_a11y}{d}}    &  {\tiny \textcolor{white}{no}}   \\
\hline
LastPass  &  {\tiny \textcolor{white}{no}}    &  {\tiny \textcolor{white}{no}}    &   {\tiny \textcolor{white}{no}}   &   {\tiny \textcolor{white}{no}}   & {\tiny \textcolor{white}{no}}  &   {\tiny \textcolor{white}{no}}   &  {\tiny \textcolor{white}{no}}    &   {\tiny \textcolor{white}{no}}   &   {\tiny \textcolor{white}{no}}   &  {\tiny \textcolor{white}{no}}    &   {\tiny \textcolor{white}{no}}   &  {\tiny \textcolor{white}{no}}    &  {\tiny \textcolor{white}{no}}    &  {\tiny \textcolor{white}{no}}    &   {\tiny \textcolor{white}{no}}   &   {\tiny \textcolor{white}{no}}   &  {\tiny \textcolor{white}{no}}    &   {\tiny \textcolor{white}{no}}   &   {\tiny \textcolor{white}{no}}   &  {\tiny \textcolor{white}{no}}    &   {\tiny \textcolor{white}{no}}   &  {\tiny \textcolor{white}{no}}    &  {\tiny \textcolor{white}{no}}    &  {\tiny \textcolor{white}{no}}    &   {\tiny \textcolor{white}{no}}   &  {\tiny \textcolor{white}{no}}    &  {\tiny \textcolor{white}{no}}    &  {\tiny \textcolor{white}{no}}    &  {\tiny \textcolor{white}{no}}    &  \cellcolor{blue_a11y}{\tiny \textcolor{blue_a11y}{d}}   &   {\tiny \textcolor{white}{no}}   &  {\tiny \textcolor{white}{no}}    &  {\tiny \textcolor{white}{no}}    \\ \hline
McAfee  &  {\tiny \textcolor{white}{no}}    &  {\tiny \textcolor{white}{no}}    &  {\tiny \textcolor{white}{no}}    &  {\tiny \textcolor{white}{no}}    & {\tiny \textcolor{white}{no}}  &  {\tiny \textcolor{white}{no}}    &   {\tiny \textcolor{white}{no}}   &   {\tiny \textcolor{white}{no}}   &  {\tiny \textcolor{white}{no}}    &  {\tiny \textcolor{white}{no}}    &  {\tiny \textcolor{white}{no}}    &  {\tiny \textcolor{white}{no}}    &  {\tiny \textcolor{white}{no}}    &  {\tiny \textcolor{white}{no}}    &    {\tiny \textcolor{white}{no}}  &  {\tiny \textcolor{white}{no}}    &   {\tiny \textcolor{white}{no}}   &  {\tiny \textcolor{white}{no}}    &  {\tiny \textcolor{white}{no}}    &  {\tiny \textcolor{white}{no}}    &   {\tiny \textcolor{white}{no}}   &  {\tiny \textcolor{white}{no}}    &  {\tiny \textcolor{white}{no}}    &  {\tiny \textcolor{white}{no}}    &  {\tiny \textcolor{white}{no}}    &  {\tiny \textcolor{white}{no}}    &  {\tiny \textcolor{white}{no}}    &  {\tiny \textcolor{white}{no}}    &   {\tiny \textcolor{white}{no}}   &  {\tiny \textcolor{white}{no}}    &   {\tiny \textcolor{white}{no}}   &  {\tiny \textcolor{white}{no}}    & \cellcolor{yellow_a11y}{\tiny \textcolor{yellow_a11y}{i}}    \\
\hline
\end{tabular}
\vspace{0.1cm}
\caption*{\textcolor{blue_a11y}{\rule{0.3cm}{0.3cm}} Deliberate uptake \hspace{0.3cm}
  \textcolor{yellow_a11y}{\rule{0.3cm}{0.3cm}} Impromptu uptake}
\end{table*}

\subsubsection{\textbf{How BLV Users Start Using PMs}}
\hfill\smallskip
\label{sec:uptake}

\noindent We identified two distinct paths users take when initially engaging with a PM. During these initial steps of an adoption process, users can be exceptionally vulnerable to system failures (often caused by accessibility barriers), which would immediately stop their efforts.
If they utilize multiple PMs, users can employ both strategies at different stages of their journey, sometimes even contemporaneously.

\paragraph{Deliberate uptake.} Participants often started using a PM after receiving recommendations from trusted people. Those could be partners, such as for P6: \inquote{My partner said ‘Hey, I really love 1Password. You should try it.' And that's when I started using it.} But also parents, close friends, or teachers advocated for PMs, as P15 described: \inquote{One of my friends was singing the praises of 1Password to me. I'm like, all right. We can give this a shot. This sounds like the solution to my problems.} Other participants started looking into PMs on their own initiative, for instance P17:  

\begin{quote}
    \inquote{I had discovered [PMs], and then I was like: ‘Oh, I should maybe keep my password safe.' Also I was starting to create a lot more accounts, especially during high school. I needed to have actual accounts for things as opposed to just using them as guest. So I just decided to figure out a little way to keep my passwords organized, instead of just having a bunch of Post-it notes everywhere.}
\end{quote}

Finally, a few participants turned to PMs after suffering a security breach of one of their accounts, as was the case for P21: \inquote{That's when I really started using the iPhone created passwords. Because I thought to myself, well, you obviously failed with the password you had. So let your phone do something random that no one can guess.} 
What they all share is a desire to improve their security practices. This motivation is often rooted in the awareness that their current password management strategies are not ideal. Consequentially, users believe PMs enhance their accounts' protection. P15 explained: \inquote{Before I used 1Password, I would just kind of use a password variance scheme where I would have the same root password, and I would change like one element of it, depending on the kind of service it was. So I was moderately unsafe to be clear, like I was repeating passwords, but I was repeating them in a kind of organized manner that made sense to me, and hopefully no one else.}

Furthermore, a deliberate uptake was often linked to a single, substantial setup effort. This includes picking out a PM, getting it up and running on their system, and transferring the accounts over. Some users even went as far as replacing predictable passwords with randomly-generated ones in the process. Participants described this process as cumbersome. Given sufficient motivation, users would still push through, as exemplified by P3: \inquote{At first it was a bit of a headache, because you had to basically put in each password one by one. It was annoying, because nowadays we have so many accounts. [...] And I ended up realizing that some of the passwords I hadn't updated in my offline format. So I had to click forgot password, generate new passwords for all of them, and then put them in that way. But once that was set up, things flowed way more smoothly the more passwords I got in.} For other users, however, the required effort proved to be too steep of a hurdle, as P19 described: \inquote{My partner uses LastPass and I wanna try to use it. But I keep forgetting to sit down and figure out how to do it, because I have so many passwords in Apple Keychain. And I don't know how to transfer all of them to LastPass, or even if you can do that, so that would just mean manually entering in all of the passwords. [...] That just is like a big task that I don't want to do.}

This uptake strategy closely matches findings from previous work on sighted users' journeys towards adopting a PM; both BLV and sighted users were motivated by a desire to improve their accounts' security~\cite{Zibaei2023simple-nudgepassmgnrs, pearman-19-use-pw-manager, ray-21-pw-manager-old}, especially so if they suffered a security breach~\cite{alkaldi-16-adoption-phone-managers, pearman-19-use-pw-manager}, and word-of-mouth proved to play a crucial role, especially if recommendations came from closely trusted people~\cite{alkaldi_encouraging_2019, mayer-22-pwm-edu, oesch-22-started-pw-manager}.

\paragraph{Impromptu uptake.} Besides a deliberate uptake, participants often started using a PM simply \inquote{because it was there}(P3). This often involved receiving a notification from a preexisting PM on their phone or computer or built into their browser. While logging into an account, users would be prompted whether they want to have their credentials saved. Similarly, in the process of registering a new account with an application or website, the system or browser would suggest a randomly generated password to users, which would then be directly stored by the PM. Since these prompts were largely accessible and picked up easily by screen readers, participants frequently chose to agree. While most had an implicit understanding of the underlying mechanisms in place, some did not realize they had just interacted with a PM. P8 exemplifies such unknowing usage:

\begin{quote}
    \inquote{I just did an iOS update, and mysteriously there is something called ‘Passwords' that's popped up. And when I open it up, it just recognizes my face and I see passwords from [my] old cell phone number. And I had linked accounts like Gmail [etc.] and it's weird how I haven't been active in that for a couple of years, and now, all of a sudden, on this ‘Passwords' app it shows all my old passwords.}
\end{quote}

Following this adoption pattern, users gradually add accounts to their PM over time, resulting in a more organic and effortless uptake. Some started looking more into this technology, discover new features, and slowly replace their old password with newly generated ones. P23 explained: \inquote{I periodically will sit down and update my password, especially now that I have 1Password. For the last couple of years I've been slowly working my way through updating.}

Previous studies found that sighted users often follow a similar progression when adopting PMs \cite{Munyendo2023switchingpassmngrs, oesch-22-started-pw-manager, amft-23-cred-management, pearman-19-use-pw-manager}. This suggests that the nudging mechanisms to store credentials are accessible and provide comparable benefits to sighted and BLV users.

\subsubsection{\textbf{The PMs BLV Users Use}}
\hfill\smallskip

\noindent Participants used a variety of PMs, often even multiple in parallel. Some had specific reasons for doing so, such as wanting to separate work from private life or using different PMs on different devices. Others impromptu adopted a second or third PM or deliberately wanted to get a better-suited solution after organically taking up their first PM.

\paragraph{Third-party.}
Third-party PMs are standalone applications not integrated into an operating system or a browser. These tools usually need to be installed manually.
Among these PMs, participants preferred 1Password, which they recognized as being highly accessible. It was frequently recommended by one BLV user to another. In general, participants started using third-party PMs out of a deliberate motivation. One exception to this was P33, who started using a PM organically after installing a security suite for antivirus and firewall software, which happened to ship with a PM. Participants had mixed conceptions around the security and accessibility guarantees of third-party PMs. Some believe that third-party developers might have fewer means and incentives to make their systems accessible. Also, users hesitated to trust vendors they did not already know when it comes to security. P16 stated: \inquote{I'm less confident with smaller third-party companies, where they might not be aware of accessibility. And even if they are, maybe there's less motivation to work on it.}
On the contrary, other participants expected better security and accessibility from third-party tools, since these often charged a fee, as opposed to the free built-in solutions. P1 explained: \inquote{The third-party apps are more secure because that's the objective of the PM company, to run that PM and be able to focus on that.} Another reason for choosing a third-party PM was cross-platform capabilities, which built-in solutions often lacked.

\paragraph{OS built-in.}
Some operating systems offer built-in PMs. Especially on mobile phones, participants often turned to these solutions, both as part of a deliberate or impromptu uptake. Since most of the participants used an iPhone, the integrated Keychain software was the most commonly used PM. Participants valued the accessibility benefits that came with built-in solutions, owing to the tight integration of screen reader and PM from a single vendor. As the vendors behind OS-based tools were mostly large, well-known enterprises, participants had extended trust in the software when it comes to security. P4 stated: \inquote{I think Apple's implementation of their PM is probably on the more secure side, just based on the way Apple does things.} Cross-platform compatibility issues often prevented participants (who mostly ran Windows on their desktop) from exclusively using an OS built-in solution, as these where rarely compatible with systems from different vendors.

\paragraph{Browser built-in.}
Finally, participants stored passwords in browsers, most commonly Google Chrome. Most often, the usage began impromptu with a notification from the system, prompting the user to save their credentials after logging in or suggesting a generated password when creating a new account. Beyond these immediate cues, participants perceived browser-based tools as nontransparent and difficult to explore, due to the obscure integration into the browser. This missing transparency also had users question the security of browser-based PMs. The direct association between these tools and the internet further lowered users' trust, as P4 highlights: \inquote{Browser-based ones are probably the least secure, mostly because there are zillions of exploits for browsers, and there's always some sort of cross-site scripting thing going on.}

Overall, when it comes to choosing a PM, previous work showed that users highly rate the perceived usefulness~\cite{Zibaei2023simple-nudgepassmgnrs,Munyendo2023switchingpassmngrs,stobert-14-pw-life-cycle}, and our participants where no different. In contrast to sighted users, however, usefulness for BLV participants was tightly linked to accessibility, making it the decisive factor for adoption.

\subsubsection{\textbf{Effective Feature Usage}}
\hfill\smallskip

\noindent PMs can only offer strong security when multiple features work together seamlessly.
All of the participants used PMs to store (some) of their passwords, and most retrieved them via the autofill feature. A few, however, relied on copying and pasting passwords.
This latter practice can undermine the security of PMs, since it cancels the built-in phishing protection that comes with autofilling passwords~\cite{silver-14-pw-manager}. Furthermore, the clipboard retains its contents for a predefined time, which can leak passwords to other users on the same system~\cite{fahl-13-clipboard-pw-manager}.

Another substantial security benefit of PMs is the ability to create strong random passwords~\cite{Zibaei2023simple-nudgepassmgnrs}.
Some participants did utilize this feature, and most of them recognized its security benefits. P21 stated: \inquote{I've started to use it more now than I used to, because I think my passwords were more easily guessed, and I just feel like this is more random and harder for someone to figure out. So I do tend to use that feature a lot now.} 
In terms of accessibility, participants' perceptions were divided. 
Heavy users of the autofill feature appreciate the usability and convenience.
Accessible keyboard shortcuts proved to be vital for this positive experience. 
P22 pointed out:

\begin{quote}
    \inquote{I find it  very easy that it auto populates things for me, just push this button, and you can log into your account. I don't have to think about what my password is, and it also feels secure because it's been generated with these random strings of letters and numbers and special characters that people aren't gonna try to guess organically.}
\end{quote}

On the contrary, some users struggled with synchronizing randomly generated passwords across the various systems they were using. If they did not use the same PM on all devices, or password synchronization was not working for them, participants occasionally had to type in passwords  manually. In this case, users expressed difficulties locating passwords, navigating through them character by character, or entering rarely used special characters on their keyboard. Therefore, users would reject using generated passwords extensively, as highlighted by P23: \inquote{1Password lets you generate their passwords. But I'm always a little nervous to do that, because visually, it's kind of hard to go back and forth if I need to type it in.}

Finally, some PMs inform users about their passwords showing up in a leak or generally being easily guessable. Several participants came across this feature and appreciated the insights it provided. P6 explained:

\begin{quote}
    \inquote{I love that it has this thing called watchtower, and it will tell you if something has been compromised. It has a list of your passwords that have been found on the dark web and you can go change them. [...] So I appreciate those little hints that help you keep all your accounts updated.}
\end{quote}

Overall, while PM usage was widespread among participants, they rarely used the entire feature set and, therefore, missed out on several security benefits. Since previous work found similar patterns among sighted users~\cite{pearman-19-use-pw-manager,ray-21-pw-manager-old,mayer-22-pwm-edu}, this suggests that barriers exist beyond input mechanisms and across communities. In the following, we will revise the role of accessibility in this context.

\subsubsection{\textbf{Navigating PM Accessibility}}
\hfill\smallskip

\noindent Since a technical system's accessibility is indispensable for BLV users, we put the spotlight on this aspect of PMs.
A common concern among participants was whether a system they hadn't used before would be accessible to them. 
Users felt hesitant to try out new technologies for critical purposes, such as managing passwords. This sentiment was rooted in past experiences with inaccessible software, as P10 pointed out: \inquote{There are loads of apps that are not fully accessible, for VoiceOver especially.} Also, systems could start out being accessible but then deteriorate over their lifespan, as P31 highlighted: \inquote{Like so many times, software will be accessible, and then they'll forget about accessibility and deprioritize it.} However, since all of the participants employed PMs, we concluded they must have managed to overcome such concerns. We dug deeper into this and identified three avenues, depending mostly on the initial uptake experience. 

\paragraph{BLV recommendations.}
As previously described, recommendations from trusted people are a major factor for a \textit{deliberate uptake} of PMs. We recognized that, if these recommendations came from fellow BLV users, they could effectively alleviate accessibility concerns. P27 provides an example: 

\begin{quote}
    \inquote{I was deeply concerned about [accessibility]. But I've had other acquaintances and friends who have used it for a number of years, and they have been, you know, it's worked for them. And so it took me a while to get my toe in the water just because of the accessibility concerns but then, once I did, I felt like it was pretty easy swimming.}
\end{quote}

\paragraph{Trusted vendors.}
Some users were willing to give certain vendors the benefit of the doubt when it comes to accessibility standards. This was the case specifically for Apple Keychain. Participants had made many positive experiences in the past, which established their trust in novel software also being accessible. P21 stated: \inquote{I know that for the most part iOS is pretty accessible. So when I found out they had a PM, I figured, if everything else is accessible, hopefully, this is [too].}

\paragraph{Accumulated trust.}
Finally, users who adopted a PM impromptu did not experience accessibility concerns in the same way. Using the system for the first time was mostly a spontaneous decision, and the initial prompt was accessible to those who chose to store their password with the system. Hence, users put little thought into long-term accessibility concerns. We found that they established trust through usage over time. As they encountered no roadblocks, users gradually extended their usage of the PM. P7 described slowly warming up to using a PM: \inquote{In the beginning I was clicking on ‘No, not now' and then, after a while, I thought, I will give it a try, and then it works. And then I kept using it for a couple of other [passwords].}

Overall, we found that participants perceived the PMs they are using to be easily accessible. P12 said \inquote{I think what I have now meets my needs in terms of accessibility.} Similarly P9 stated: \inquote{Overall, just the the whole Apple [password] manager is really good. There's really no issue with it. It's pretty flawless.} Sometimes, PMs even exceeded users' expectations, as P3 exemplified: \inquote{That's one thing that surprised me, accessibility has been pretty seamless.}

\subsubsection{\textbf{A PM is a Convenience Tool}}
\hfill\smallskip

\noindent We identified that users stick with PMs predominantly for convenience reasons. They enjoyed how PMs eased their daily workflows. P23 noticed: \inquote{It's gonna save me 30 seconds, up to a minute. Having [passwords] in Chrome is very convenient.} Convenience, while often not the primary reason to get started on a PM, became a crucial factor for long-term adoption. P5 explained: \inquote{I find them to be really convenient. I mean, I think that's how they should be, I feel like a PM is like working for you.} Occasionally, the convenience could even trump over security benefits in the users' perceptions. P14 stated: \inquote{It's very convenient. It's very secured. If you do it right. It is great, but it has to be done right. You can't do it at the cost of convenience. If you want people to use your app. make it convenient.} Among the convenience aspects of PMs, we established the following major factors:

\paragraph{Reduce mental burden.}
Participants typically kept a high number of accounts. Save for password reuse, this came with a severe mental load of remembering passwords, usernames and e-mail addresses for each account. Participants felt overwhelmed with keeping up with this ever-growing demand, which often led them to forget their passwords and having to go through a cumbersome reset process. This was particularly the case for accounts that they did not use frequently. Participants appreciated the storage feature of PMs, which reduced frustrating experiences and \inquote{relieve a lot of stress} (P1). P18 summarized this sentiment: \inquote{Not everyone has the cognitive capacity to remember passwords, or even if they write them down, they may not remember where they place them, so it just makes it easier to have an electronic reminder or something that can keep all that information for you.}

\paragraph{Get things organized.}
Another major factor for many participants was having a place to file all their accounts they set up over time. However, many also stored other important data and documents in their password vaults. In essence, PMs served as a one-stop solution that provided users with an accessible overview. Some systems allow users to label accounts, making them easier to find, as P2 noted: \inquote{I'm really thorough about sorting all of my passwords into groups or tags.} Finally, some users cherished the option to share selected passwords with family members or business partners. P1 summarized their preferences:

\begin{quote}
    \inquote{Shared vaults are really helpful. Also, being able to keep multiple types of data and not just a password. For example, the serial numbers for software that I have go into 1Password, banking details go into 1Password, card information, everything that I find to be important goes into 1Password, so I know where to find it. I can just search for it right there in 1Password.}
\end{quote}

\paragraph{Diminish manual password typing.}
Participants largely agreed that having passwords automatically inserted was a big improvement in convenience. Especially while using mobile phones, they found manually typing in passwords to be cumbersome, even more so if they included special characters. P5 expressed: \inquote{I don't wanna be typing in passwords, like, that's annoying to me.} Users also sometimes struggled with typographical errors, as highlighted by P22: \inquote{I don't have worry about typing it in wrong, because that's a thing, when you're visually impaired, is entering information in incorrectly.} Such mistakes meant users had to awkwardly try and correct the password, or even worse, failed login attempts would accumulate, which could result in the account being locked. Finally, autofilling passwords would save a substantial amount of time. P16 explained: 

\begin{quote}
    \inquote{When I'm using a Bluetooth headset, there's screen reader lag, that can impact me entering a long password. To avoid that, having Face ID or something like that just speeds up that process.}
\end{quote}

This quote also highlights how users strongly favored using biometric authentication methods to access their accounts. They described them as easy to use and accessible. Particularly fingerprint authentication worked well out-of-the-box for many users. While most users made extensive use of facial recognition on their phone, some participants failed to operate it successfully. Still, users had high trust in the security of biometric authentication. P20 stated: \inquote{I wish the fingerprint feature was still available, too, because, frankly, if I had both [Touch ID and Face ID], I would use both of them to get into my iCloud, Keychain, or any kind of account.}

Convenience was also the main driving factor for sighted users, when it comes to sticking with PMs~\cite{Seiler-Hwang2019-dontusepassmngrs}. Similar to BLV users, they enjoyed the reduced mental load~\cite{Munyendo2023switchingpassmngrs,alkaldi_encouraging_2019} and need for manual typing~\cite{mayer-22-pwm-edu, alkaldi-16-adoption-phone-managers}. For similar reasons, sighted users also favored biometric authentication~\cite{alkaldi-16-adoption-phone-managers, Seiler-Hwang2019-dontusepassmngrs}. The latter two proved to be even more relevant for BLV participants, for whom typing was even more cumbersome on mobile devices, while biometric authentication was usable without the need for assistive technology.

\subsubsection{\textbf{Authentication in Public}}
\hfill\smallskip

\noindent One working hypothesis we held at the start of this project was that authenticating in public using a screen reader was difficult, since bystanders would hear the characters being read back as the user was typing~\cite{ahmed_privacy_2015, jain_smartphone_2021}. Hence, PMs could provision the added benefit of providing biometric authentication to all applications. We questioned the participants on their experiences with this matter. We found that, while the participants were aware of the issue, they were not particularly concerned, as they routinely concealed their in- and outputs in public regardless. They relied on screen curtain to prevent shoulder surfing. Also, most participants reported generally using at least one earbud to better understand the screen reader, which would shield against eavesdropping. Even without headphones, some participants expected that sighted bystanders would not be able to comprehend their screen reader, as it was set up to talk at an accelerated pace. P5 summarized this sentiment:

\begin{quote}
    \inquote{The best part about being blind is you can have your curtain on your phone. You don't have to see your screen to know. So you only hear it with your AirPods or your headphones, so no one ever knows what's on my screen, because my screen curtain is on so. And then the speech rate on my phone is really fast, so all they can hear is, they won't even know what is going on. So that's the best part about being blind like people, they just never know. So I think that I will always win.}
\end{quote}

Still, ultimately, users valued the ability to use biometric authentication everywhere, as it substantially enhanced their convenience. This exemplifies how users can adopt secure practices even when security is not the motivating factor.

\subsection{Theme II: The Absence of \textit{Practical} Accessibility Restricts Secure Behavior}

Even though all participants used PMs, we rarely observed a full adoption based on \textit{practical accessibility}. This means that users would employ various other techniques to manage their passwords in parallel rather than using the built-in features for strong and unique password generation and for autofill.

\subsubsection{\textbf{Users Engage in Suboptimal Password Behavior}}
\hfill\smallskip

\paragraph{Password reuse.} Participants would occasionally use the same password across multiple accounts, which they regarded as low-importance or shared with others. This practice was rarely done using a single password, but it more commonly involved several passwords that users would cycle through. P17 stated: \inquote{I do have a couple of passwords that I know work and are kinda hard to guess just because of their nature. So I do reuse them.} Some participants stuck with this technique even when their password was not conforming to a system's minimum requirements, as P28 explained: \inquote{Sometimes some places need a long password, sometimes they don't, but in cases of where a password requires more characters, I just type the password twice. So, for example, if it's like 123, then I just do 123123.}

\paragraph{Predictable passwords.} A common practice among participants was generating passwords following a certain personal scheme. By combining words, numbers, and special characters in meaningful ways, the resulting passwords held a significance to the users, which only they expected to understand. In contrast to password reuse, this technique was also used for high-value accounts, such as online banking, in case participants deemed the resulting passwords highly secure.
P5 described: \inquote{So I try to think of something that I'm into like in the moment, whether it's something that I saw on Netflix, and then I try to change the lettering I put. You know I substitute some letters with numbers, like the ‘E' I turn them into number threes, the ‘I' I turn them into number ones.} This process of replacing numbers with letters is often called \textit{leetspeak}~\cite{nisenoff-23-reuse}. Users familiar with braille would sometimes use a related approach, as outlined by P7:

\begin{quote}
\inquote{For braille, one letter is usually one dot out of six,
or multiple dots out of those six in different combinations. And the way you remember that is basically the left three dots are 1, 2, 3, and then the right three are 4, 5, 6. For example, if I were to use the password ‘RUNNER123', I would write ‘RUN', and then the contraction for ‘ER' is 12456. So I would just write ‘12456', and skip the 3, and then add a separate set of numbers different to that.}
\end{quote}

Both of the above-mentioned techniques ultimately enable people to memorize their passwords, by either minimizing the number of passwords one has to remember (\textit{password reuse}) or lowering the complexity and facilitating memory recall (\textit{predictable passwords).}

\paragraph{Noting down passwords.} Participants often reported taking digital notes of passwords. These records could take the form of text files or spreadsheets on a computer, or they could be stored in a notes app on a smartphone. Some users described protecting their credential lists with a password or biometric authentication. Users would then copy and paste their login information when needed. A handful of participants tried to secure this method further by not directly noting down passwords but rather only fragments, hints, or encoded versions, as exemplified by P7: \inquote{When I'm writing them down I have a system. I don't write them down directly. I have a code. I write them down that way. So if someone finds the file, they won't be able to read it immediately.} 

\paragraph{Brailling passwords.} Similarly to noting down passwords digitally, several participants also described creating physical password lists using braille. Using either a braille embosser or a slate and stylus, users can write out their passwords on paper, which they then store in a secure location. When authenticating, they can read back their password character by character and type it in. The primary advantage of this technique is its accessibility. Users fluent in braille have no issues storing and retrieving their passwords. The main downside on the other hand is availability. The lists would not always be at hand when needed since carrying around significant amounts of paper was cumbersome. Participants also mentioned how password lists would wear out over time, get damaged or lost. Therefore, users need to constantly invest efforts into maintenance. P11 explained: \inquote{I've made it a goal for myself before the end of this year to redo my password booklet. There's this different form of braille paper that's essentially sheets of thick plastic. So rather than paper, you would type onto this plastic sheet, and that makes your braille more permanent.}
Participants were divided when it comes to the perceived security aspects of brailling passwords. Some held the opinion that writing down passwords is inherently insecure. P16 compared it to the infamous practice of \inquote{writing a password on a sticky note and just leaving it out.} On the other hand, physical lists were described as hacker-proof, due to their offline nature. More importantly, those in favor of this technique attributed its security to the perception that, only BLV people could really access it. P5 stated: \inquote{I write it out in braille so that only I can read it,} and P3 summarized their method as follows:

\begin{quote}
    \inquote{I will write the password down on an external medium like braille and I'll actually write it on paper instead of any app. You know, you write something down on paper, there's no way it's ever going to get online. There's no way that it's ever going to be accessible by somebody other than me. And because I'm writing in braille, you know the odds of someone accessing it kind of decrease even more. [...] Anything that's online can be hacked.}
\end{quote}

Ultimately, we classify this as ‘security through obscurity.' P20 explained their perception: \inquote{Unless somebody knows braille and they route through my desk, they're not gonna be able to find it. [...] How many blind people go into professional robbery, into other blind people's houses, even if, they might not even know I'm blind.} 

When comparing these findings to previous work on sighted users, BLV users engage in much the same suboptimal password behavior of reusing passwords~\cite{pearman-19-use-pw-manager,Munyendo2023switchingpassmngrs,mayer-22-pwm-edu}, choosing predictable passwords~\cite{stobert-14-pw-life-cycle,pearman-19-use-pw-manager}, and writing down passwords~\cite{stobert-14-pw-life-cycle,mayer-22-pwm-edu}. Specific to our study's participants is the use of Braille, both to keep physical records, and as password alternation strategy.

\subsubsection{\textbf{Barriers for Secure Password Behavior}}
\hfill\smallskip

We dug deeper into the rationale behind engaging in practically less secure password management techniques while also employing PMs. We found two main contributing factors:

\paragraph{Need for autonomy.} The first reasoning follows the theory of self-determination~\cite{ryan_self-determination_2000}, revolving around the users' desire for autonomy. According to Alkaldi and Renaud~\cite{alkaldi-16-adoption-phone-managers}, \textit{autonomy} is defined as \inquote{the sense of freedom and control over ones own choices.} By relying entirely on a PM, participants felt like having little control over their passwords. Automatically generated passwords could undermine a user's sense of agency, as they would be difficult to remember or even comprehend. P25 stated: \inquote{I don't like not knowing what the password is. So I'd never do the suggested password, where it's like a bunch of letters and numbers. I like knowing what it is and not having to search it through the PM, that confuses me.} Even using autofill instead of manually typing in a password can evoke similar feelings, as highlighted by P21: \inquote{[For my] bank, I always enter the password. I never trust having that be auto-populated or anything like that.} This quote also showcases how this sentiment is even more distinct for high-value accounts, which is why users often choose to exclude such accounts from their PM. 
To counter this desire for control and agency, and to satisfy their need for enhanced security for high-value accounts, users turn to highly reliable and trustworthy methods. This could simply mean using a subjectively stronger scheme to come up with a password for such accounts, not reusing a password anywhere else, or choosing a subjectively more secure PM for this specific job, as exemplified by P23: 
\begin{quote}
    \inquote{if it's not a financial [account], or what I consider to be one I really want to protect, then Google Chrome will fill it in for me. If it's financial, then I'm gonna copy it over from 1Password.}
\end{quote}
However, we also encountered users who specifically refrained from saving high-value accounts in their PM. P3 explained: \inquote{I'm not [storing] passwords that are, you know, do or die kinds of passwords like my financials,} and P4 resorted to memorizing passwords as a means to maintain control: \inquote{I have two accounts that aren't in [my PM]. I know those passwords. And I don't really wanna change it to something that I don't know, cause I don't know any of my passwords that are in 1Password.}

In essence, participants acknowledged that fully committing to a PM would create a strong dependency, which could go against their demand for independence. This sentiment is also present in sighted users~\cite{alkaldi-16-adoption-phone-managers,zibaei-22-random,pearman-19-use-pw-manager,Seiler-Hwang2019-dontusepassmngrs}, but appears to be even more deeply rooted in BLV participants who despise being dependent on yet another (technical) system.

\paragraph{Dependence on fragile system.} A strong dependence on the system can cause significant problems for users in cases when the PMs fail to function correctly. This could happen due to bugs in the software. However, even benign updates often compromised the systems' accessibility. P1 explained how upgrading their PM to the latest version came at the expense of accessibility: 
\begin{quote}
   \inquote{1Password started out with unlabeled buttons when they switched to Version 8, which made me as a consumer feel like, ‘Oh, we're switching to version 8. But we're not asking our subset of users who are blind using access technology for feedback on this software'.}  
\end{quote}
This was a common experience for BLV users and could often steer them away from adopting PMs, as P27 highlights: \inquote{My concern was: I'd get on board with this system. I put in all this effort into learning how to make it work, and then, you know, some code monkey would change something, and all of a sudden it wouldn't work with JAWS or NVDA.} This problem extended beyond the PM itself: the malfunctions might also be attributed to the primary systems users aim to access, as P12 pointed out: \inquote{If an app is updated, or if an iOS update happens, and then something doesn't work, then you don't have access to [your account] potentially, because that seems to happen a lot.}
Participants, therefore, turn to fallback mechanisms to offset such system failures. It is crucial for these backup solutions to be highly accessible to avoid being exposed to the same risks. Meanwhile, convenience is less imperative, as backup solutions are expected to be used rarely. P5 explained: \inquote{You should always have a backup plan, because you never know, technology will fail you at the moment that you need it the most. So I don't dismiss the fact that I have it written down, but obviously having it in your PM is going to be much more efficient and conducive.}

\paragraph{Underlying trust issues.} Both of the aforementioned factors indicate that users still hold trust issues in regards to PMs. These sentiments can be rooted in skepticism of technology in general, as showcased by P13: \inquote{I don't know, but I don't always trust that the technology is going to do what it's supposed to do and and so I always want to make sure that [my password] is also something I can remember, which, probably, is bad.} 
Also a lack of technical understanding, which diminishes the feeling of being in control, negatively affects users trust, as explained by P25: 

\begin{quote}
    \inquote{I feel like in the back of my head, similar to why I don't want [PMs] to use random passwords, I have trust in that they're going to keep them, but not enough trust where, like, I don't know the ins and outs of how it works, whereas, my own system seems very straightforward.}
\end{quote}

In accordance with findings related to sighted users~\cite{alkaldi-16-adoption-phone-managers,Seiler-Hwang2019-dontusepassmngrs,amft-23-cred-management}, we identified trust issues throughout most interviews. Additionally, we recognized ways to establish trust between participants and PMs. If they engaged with a PM over an extended time, they would slowly build up trust towards that system. In general, if users felt informed and oriented about their PM's security practices, they demonstrated higher trust. Therefore, a certain openness in both the vendor's security practices and the PM's functionality were crucial. P9 highlights: \inquote{I would say with the Apple one, I'm pretty confident.[...] The Google one, I'm less confident in. Because I don't know how to access [my passwords], and then I just don't have much experience with that, so I can't trust it.}
Finally, even convenience could foster trust in the long run. This added benefit could alleviate initial concerns and, thereby, kickstart trust building. P33 explained:

\begin{quote}
    \inquote{I believe [my PMs] are as secure as they can be, or else I won't put my passwords in there, because for the longest time with McAfee, I didn't allow it to save passwords. But then I realized how this is much quicker, so just let me do it this way.}
\end{quote}

\section{Discussion} \label{sec:discussion}

We relate our findings to the concept of \textit{positive security,} examin the security implications of braille password lists, provide implications for design and suggestions for developers, and discuss the role of PMs in a passwordless future. Finally, we provide ethical considerations around our research and recognize its limitations.

\subsection{Ontological and Positive Security}
The underlying problems of trust, negative accessibility experiences, dependence on fragile systems, and lack of autonomy are examples of deeper issues relative to BLV individuals' \textit{ontological security} and \textit{positive security} (borrowing from the analytical lens in the context of \textit{cyber}security introduced by McClearn et al.~\cite{McClearn2023}). According to the sociologist Bill McSweeny, ontological security considers the freedom to live free from fear and protection from harm~\cite{mcsweeney1999security}. Ontological security is seen as ``security as being'' rather than ``security as survival''~\cite{Steele2025, Mitzen2006}. Croft contextualizes ontological security as creating a sense of security through leveraging trusted relations and routines~\cite{Croft2017}. A closely related concept is Roe's notion of positive security as the ability to pursue one's interests and fulfill one's needs through trusted relations~\cite{Roe2008}. Positive security focuses more on the ability to withstand and navigate (cyber) threats rather than on the dominant, negative security perspective that emphasizes the need for protection from threats (in cyberspace)~\cite{Renaud-Coles-Kemp}. 

When our results are seen through these lenses, it becomes apparent that PMs do not align with the BLV users' ontological security. These tools fail to create a sociotechnical cybersecurity model that conceptualizes practical accessibility as a benefit and protection~\cite{giddens2023modernity}. This precludes PMs from engaging with the agency of BLV users as stakeholders and their way of establishing ontological security. The evidence of reusing passwords or keeping high-value passwords outside of PMs, taken together with the negative experiences with fragile systems that never prioritize accessibility, preclude BLV users from the positive security feeling that their understanding of, and position within, cyberspace is stable and reliable~\cite{mcsweeney1999security, Roe2008}. The ontological insecurity demonstrated through brailing, fast screen reader output, and reliance on screen curtains as ‘security through obscurity' does little to cater to the confidence essential for BLV users ‘security as being'. Even though they know that these practices are not secure, they use them anyway, as password management for them is a simple matter of ``security as survival.''

\subsection{Security Implications of Brailling Passwords}

We found that some users revert to braille as a means of keeping physical password lists. While these have clear accessibility advantages and can serve as a viable backup solution in case of a technological outage, we want to discuss potential security pitfalls of this mitigation technique.
Braille password lists are susceptible to physical deterioration, where even a single altered dot can change the validity of a password. Hence, these lists need to be constantly redone, as some participants described. 

More importantly, a few participants described how braille password lists are more secure than written text, as a sighted attacker would not be able to read them. While it is true that a casual drive-by attacker (e.g., a burglar) might not immediately recognize a password list in braille, this does not protect against targeted attackers, unless the list is stored in a secure location. Ultimately, it can be described as ‘security through obscurity.' The same is true for the technique of replacing characters by their braille number representation when creating a password. While these alterations might not be covered by the most common password-cracking approaches, once known, they become trivial to bypass. Thus, these techniques can leave BLV users with a false sense of security.

\subsection{Implications for Design}

Based on our findings, we derive implications for the design and development of future PMs:

\paragraph{Adapt password generators to BLV needs} 
Both 1Password and Keychain treat hyphenation in password generation as a flexible option for improving memorability, typically by separating words. This resemblance to phone number chunking caters to sighted individuals' short-term memory. However, it does not consider the actual workings of various screen readers which navigate passwords character by character and pronounce the ``-'' character as dash, minus, or hyphen, which can cause confusion. Moreover, obscure special characters can be difficult and time-consuming for BLV users to located on a keyboard.
Therefore, we recommend generating passwords with unambiguous, commonly used special characters that are easily navigable by screen readers. Passphrases could be pronounced word-by-word rather then character-by-character.
Furthermore, password generators could leverage BLV individuals’ superior memory performance -- particularly in areas such as digit span forward, name learning, and word span~\cite{withagen2013short} -- as a potential basis for strong, accessible recommendations.

\paragraph{Use biometrics whenever possible}
Biometric authentication, especially fingerprint authentication, has proven to be the preferred authentication method for many BLV users. It provides intuitive convenience benefits without the need for additional assistive technology.
PMs should, therefore, offer it whenever possible and actively guide users to setting up this authentication method during the early stages of adoption.
This is particularly relevant for impromptu uptake scenarios as prompting the user to set up biometric authentication for their newly saved password could also help raise awareness of the PM they just interacted with.

\paragraph{Leverage trusted parties}
When it comes to evaluating the accessibility of a new tool, users often relied on fellow BLV people's recommendations or vendors who have proven to deliver accessible products before. 
These trusted parties should be leveraged by providing specific testimonies or reviews from BLV users in adequate form, e.g., audio messages. These can help alleviate accessibility concerns during a deliberate uptake of a new PM.
PM vendors can also highlight other software they developed that are well known for being accessible.

\paragraph{Satisfy need for autonomy}
BLV users displayed a strong need for autonomy and control when it comes to managing their passwords. PMs often hide stored passwords away from the users or make it difficult to view and remember automatically generated ones.
As users want to know their passwords, automatically generated passwords should be displayed in an accessible way immediately upon creation, while all stored passwords should be easily accessible and navigable for screen readers and other assistive technologies. Users should have the option to sort and arrange their passwords to have a sense of ownership. PMs can also serve as an accessible way to manage online accounts by providing a navigable overview of all the accounts a user has.

\paragraph{Make updates robust to accessibility issues}
Participants reported that software updates often had a negative impact on a system's accessibility. When it comes to PMs, such outages can prove to be devastating. To reassure users and counter their need for an accessible, but potentially insecure, backup strategy, updates should allow to easily revert back to a stable version, e.g., by providing a demo mode after which users can choose to keep the new version or not.

\subsection{Passwordless Alternatives}
Passkeys and other passwordless approaches have been adopted by high-target platforms such as Google and PayPal. However, prior work identified several barriers that prevent smaller websites from following suit, including fallback challenges, technical limitations, and costs~\cite{lassak-24-webauthn-obstacles}.
In contrast, PMs offer immediate and tangible benefits, particularly for BLV users, for whom visual cues are often inaccessible.
Autofilling credentials reduces the risk of input errors, and domain binding provides valuable phishing protection. 
When used as intended, PMs also mitigate common issues such as weak or reused passwords~\cite{lyastani-18-pw-manager}.

Passkey nudges are often presented in a similar way to automatically generated password and autofill prompts. Therefore, users could similarly adopt the technology via an impromptu uptake~\cite{lassak-21-webauthn-misconceptions}.
FIDO2 hardware keys can cater to BLV users' need for autonomy, as their physical nature makes them both tangible while also being usable without the need for assistive technology.
Finally, passworldless authentication is heavily centered around biometrics~\cite{lyastani-20-kingslayer}, which we found to be the preferred authentication method.

However, PMs will remain relevant as they are increasingly used to store and manage passkeys as well~\cite{birgisson-22-google-passkey-e2e}. 
With their support for biometric unlock and a consistent user interface across services, they are a practical tool that supports both legacy password-based and modern passwordless authentication systems for BLV users.

\subsection{Ethical Considerations}
Reporting on password management of BLV individuals as an at-risk population is not without risk of harm. While we did not record any passwords or ask for passwords to be revealed, we nonetheless report on practices that might be insecure and could be used against BLV users -- from predictable password strategies, use of passwords in braille, having them noted down in separate files, to authenticating in public. We worked with each participant to address these potential threats in a way that fits their sense of security and agency rather than just giving them advice on secure practices. When reporting personal accounts, there is always a risk of misinterpreting the results or creating misconceptions about BLV individuals' disposition to technology~\cite{Shinohara2011}. We ensured our participants that their voices will be heard regarding the elimination of insecurities and that our joint advocacy for practical accessibility is not about functionally eliminating a visual disability, nor is it evidence that BLV individuals would be helpless when it comes to secure management of passwords.

\subsection{Limitations}
Our sample was skewed towards BLV users relying on screen readers. The results might not pertain to user relying on magnifiers, large text, screen contrast, or other forms of assistive technology to access PMs. We worked with English-speaking BLV users from the US, and their lived experiences might differ from others across the world. Our findings are limited to the current password creation policies, the features offered for managing passwords, and the state of cross-platform, cross-device support. Though we left our participants sufficient time and support to express their lived experiences of managing passwords, this might nonetheless have been insufficient for them to formulate more informed and detailed responses.

\section{Conclusion}\label{sec:conclusion}

We interviewed 33 participants who relied on assistive technologies such as screen readers to understand how BLV users manage their passwords. We conducted a qualitative analysis to investigate the role that PMs play and how well they align with the needs of BLV users.
Our findings show that PMs are widely used thanks to being convenient and accessible on a technical level. 
We identified two potential paths for the uptake of PMs: 
While users deliberately picked out standalone tools, they often started adopting OS- and browser-integrated PMs after being prompted during a login process.
However, users still resort to suboptimal password behavior such as choosing easily guessable passwords or writing them down using braille. This behavior was primarily caused by trust issues, the users' aversion to over-reliance on fragile systems, and an unmet need for self-efficacy.
Security features, such as password generators, are not well-adapted to screen readers, hindering their effectiveness and adoption.
More research is needed to align PMs to the specific needs and strengths of BLV users, for which the theory of \textit{positive security} can serve as a point of reference.

\section*{Acknowledgments}

We thank all reviewers and our shepherd for their constructive feedback in the process of improving this paper. 
We express our gratitude to all interview participants for their time and valuable thoughts.
Furthermore, we thank Aziz Zeidieh for reviewing our methodology and results from a blind researcher's perspective.
Alexander Ponticello and Simon Anell are part of the Saarbr\"ucken Graduate School of Computer Science, Saarland University.

\bibliographystyle{ACM-Reference-Format}
\balance
\bibliography{acm-ccs-2025}

\end{document}